\documentclass{emulateapj}

\newcommand{\erg}{\mbox{$\rm\,erg$}}
\newcommand{\kev}{\mbox{$\rm\,keV$}}
\newcommand{\cm}{\mbox{$\rm\,cm$}}
\newcommand{\ev}{\mbox{$\rm\,eV$}}

\newcommand{\s}{\mbox{$\rm\,s$}}
\newcommand{\ks}{\mbox{$\rm\,ks$}}

\newcommand{\beq}{\begin{equation}}
\newcommand{\eeq}{\end{equation}}
\newcommand{\vv}[1]{\mathbf{#1}}
\newcommand{\lxgas}{L_{X,{\rm gas}}}
\newcommand{\lxtx}{$\lxgas$--$T_X$}
\shorttitle{X-Ray Gas Fundamental Plane}
\shortauthors{Diehl \& Statler}

\begin{document}

\title{A Fundamental Plane Relation for the X-Ray Gas
in Normal Elliptical Galaxies}

\author{Steven Diehl and Thomas S. Statler}
\affil{Astrophysical Institute, Department of Physics and Astronomy,
251B Clippinger Research Laboratories, Ohio University, Athens, OH 45701, USA}
\email{diehl@helios.phy.ohiou.edu, statler@ohio.edu}

\begin{abstract}
We report on the discovery of a new correlation between global parameters
of the hot interstellar gas in elliptical galaxies. We reanalyze archival
{\it Chandra\/} data for 30 normal early-type systems, removing the
contributions of resolved and unresolved point sources to reveal the X-ray
morphology of the hot gas. We determine the half-light radius, $R_X$, and
the mean surface brightness, $I_X$, from the gas surface brightness profiles.
A spectral analysis determines the temperature, $T_X$, of the gas within
3 optical effective radii. We find that the galaxies lie
on an X-ray Gas Fundamental Plane (XGFP) of the form $T_X \propto
R_X^{0.28} I_X^{0.22}$. This is close to, but distinct from, a simple
luminosity-temperature relation. The intrinsic width of the XGFP is only
$0.07\,$dex, nearly identical to that of the stellar (optical) fundamental
plane (SFP). This is surprising since X-ray gas masses are typically
$\sim 10^{-2}$ of the stellar masses. We show that the XGFP is not a simple
consequence of the virial theorem or hydrostatic equilibrium, and that it
is essentially independent of the SFP. The XGFP thus
represents a genuinely new constraint on the hydrodynamical evolution of
elliptical galaxies. 
\end{abstract}

\keywords{galaxies: cooling flows---galaxies: elliptical and lenticular,
cD---galaxies: ISM---X-rays: galaxies---X-rays: ISM}

\section{Introduction\label{s.introduction}}

Elliptical galaxies lie on a two-dimensional locus, known as the
fundamental plane (FP), in the space defined by the optical effective
radius $R_e$, optical surface brightness $I_e$, and stellar velocity
dispersion $\sigma_0$ \citep{DDFP,DresslerFP}. The FP is understood to
be a consequence of the virial theorem, modified by stellar population
variations and structural nonhomology for systems of different
luminosity \citep[and references therein]{Trujillo}.  FP-like scaling
relations have been found for other types of systems as well,
including galaxy clusters \citep{Schaeffer}. Several authors have
pointed out ``cluster fundamental plane'' relations involving either
optical parameters alone \citep{Adami} or mixtures of X-ray and
optical parameters \citep{Fritsch,Miller}. Because the baryonic mass
in clusters is dominated by intracluster gas at temperatures
$\gtrsim 1\kev$, X-rays are a better tracer of
ordinary matter than starlight. \citet{Fujitaa} show the existence of
a cluster FP purely in X-ray parameters \citep[see also][]{Annis},
which they show can be consistent with simple spherical
collapse \citep{Fujitab}.

An ``X-ray fundamental plane'' for elliptical galaxies---actually a relation
between stellar dispersion, X-ray luminosity, and half-light radius---has been
suggested by \citet{Fukugita}. But in normal ellipticals,\footnote{For
our purposes, normal ellipticals are those that are not at the centers
of cluster potential wells.} the mass of X-ray gas is typically only a few
percent of the stellar mass \citep{Bregman92}. Much of the X-ray flux comes
from low-mass X-ray binaries (LMXBs); if not accounted for, this guarantees a
correlation between X-ray and optical luminosity. Normal ellipticals are
thought to lose most of their gas in supernova-driven winds \citep{Ciotti}, and
high-resolution observations using {\it Chandra\/} show that many
appear disturbed, suggesting redistribution of gas through shocks, nuclear
activity, mergers, or interaction with the intergalactic medium 
\citep[e.g ][]{Fin01, Jon01, Machacek, NGC1700}.
Thus, the visible hot gas is only a tenuous leftover of a complex
hydrodynamical history, and may be far from equilibrium.

Nonetheless, in this {\em Letter\/} we report the discovery of a fundamental
plane relation {\em for the X-ray gas alone.} We show that the X-ray Gas
Fundamental Plane (XGFP) is distinct from---but as tight as---the stellar
fundamental plane (SFP). Unlike the SFP, the XGFP is not a
simple consequence of the virial theorem. In fact, the XGFP is almost completely
decoupled from the SFP, and thus constitutes a new constraint on
the evolution of normal elliptical galaxies.

\section{Data Reduction and Analysis\label{s.data}}

\subsection{{\em Chandra\/} Archive Sample and Pipeline Reduction
\label{s.sample}}

We analyze a sample consisting of 56 E and E/S0 galaxies having
non-grating ACIS-S exposures longer than $10\ks$ in the {\it Chandra}
public archive. Brightest cluster galaxies and objects with
AGN-dominated emission are excluded.  All observations are uniformly
reprocessed using version 3.1 of the CIAO software and version 2.28 of
the calibration database.  Flares are removed by iteratively applying
a $2.5\sigma$ threshold. For the quiescent background, intervals more
than 20\% above the mean count rate are excised, to match the blank
sky background fields.  We restrict photon energies to the range
$0.3$--$5\kev$, further divided into soft ($0.3$--$1.2\kev$) and hard
($1.2$--$5.0\kev$) bands.  Monoenergetic exposure maps are created in
steps of 7 in PI ($\sim 100\ev$). An image is extracted for each
$14.6\ev$-wide PI channel, and divided by the energetically closest
exposure map to create a photon-flux-calibrated ``slice.'' The slices
are summed to produce the calibrated photon flux images. Full details
of the analysis and results, along with data products, will be
presented in future papers \citep{diehlstatlera,diehlstatlerb}.

\subsection{Isolating the Gas Emission\label{s.gas}}

Point sources are identified in each band by CIAO's {\it wavdetect}
tool.  Regions enclosing 95\% of the source flux are removed and the
holes filled with counts drawn from a Poisson distribution, whose
pixel-by-pixel expectation value is determined by adaptive
interpolation using the {\it asmooth} tool in the XMMSAS 
package. A uniform background is determined from fits to the radial
surface brightness profile, and subtracted.

To remove the contribution of unresolved point sources, we use the fact that
the hot gas and LMXBs contribute differently to the soft and the hard
bands. Let $S$ and $H$ represent the background-subtracted soft and hard
images. We can express both in terms of the unresolved point source emission
$P$, the gas emission $G$, and their respective softness ratios $\gamma$ and
$\delta$: 
\begin{eqnarray}\label{linearcombination}
  S &=& \gamma P + \delta G \\
  H &=& (1-\gamma) P + (1-\delta) G. 
\end{eqnarray}
The uncontaminated gas image is then given by
\begin{equation}\label{linearcombtwo}
  G = {1-\gamma \over \delta-\gamma}\left[ S - \left({\gamma \over 1-\gamma}
\right)\, H\right].
\end{equation}

Assuming that resolved and unresolved LMXBs share spatially independent
spectral properties, we can use the resolved sources to determine the constant
$\gamma$. We take sources between $5\arcsec$ and $5R_e$ from the center,
excluding high luminosity sources ($>200$ counts).
For systems with $>10$ sources meeting these
criteria, we fit an absorbed power law model to the combined point source
spectrum, with hydrogen column density fixed at the Galactic value for
the line of sight. Integrating the model over the soft and hard bands
yields $\gamma$. For other galaxies, we exploit the
universal nature of the LMXB spectrum \citep{Irwin03}. 
A simultaneous power-law fit to all low-luminosity
($L_X \le 5\times 10^{37}\erg\,\s^{-1}$) point sources in our
sample gives a photon index of $1.603$.
This model is used to derive $\gamma$ for the source-poor galaxies.

The coefficient $\delta$ is determined similarly, from the fit of an
APEC thermal plasma model \citep{APEC} to the hot gas emission (see \S\
\ref{s.gasparameters}). This approach assumes isothermal gas
throughout the galaxy. In case of a temperature gradient, one would
have to account for the spatial dependence of $\delta$, but this
approach is beyond the scope of this {\em Letter\/}.

\subsection{Physical Parameters for the Gas\label{s.gasparameters}}

To produce a radial profile, we adaptively bin the gas image $G$ into
circular annuli. In 12 cases 
there is insufficient signal to fit the
spatial profile. We fit the remaining profiles with S\'ersic models to
derive X-ray half-light radii $R_X$ and mean enclosed surface
brightnesses $I_X$.\footnote{Fits using $\beta$ and double-$\beta$
models are unphysical in more than half the cases, implying divergent
fluxes at large radii.} We discard 14 objects 
with $R_X$ larger than
the size of the observed field. The final sample of 30 objects has
$B$ absolute magnitudes in the range $-22.5 < M_B < -19$. Twenty-one
are group members, and 14 are brightest group galaxies \citep{LGG}. Eight of
the 14, plus 3 additional objects,
are central members of X-ray-bright groups in the GEMS survey
\citep{GEMS}. We find that up to 55\% (typically 10--30\%) of the diffuse
emission in the final sample comes from unresolved LMXBs.

The spectrum of the diffuse emission between $0.3$ and $5\kev$
is extracted from a circular region $3R_e$ in radius, excluding
resolved point source regions. We fit the spectrum using the SHERPA package,
adopting a single temperature APEC thermal plasma model for the hot
gas and a power law for the unresolved LMXBs. The normalizations of
both components, the gas temperature $T_X$, and (in most cases) gas
metallicity are allowed to vary. The redshift is set to the value
given in the Lyon--Meudon Extragalactic Database
\citep[LEDA;][]{LEDA}. For low signal-to-noise spectra, the abundances
are held fixed at the solar value. The photon index of the power law
component is determined by a simultaneous fit to the spectrum of the
lowest luminosity resolved point sources (\S\
\ref{s.gas}). Single-temperature APEC models are poor fits (reduced
$\chi^2>2$) to 12 objects. In these cases the temperature should be
interpreted as an emission weighted average. Excluding these galaxies
from the sample does not change the results.

All errors are assumed to be Gaussian, described by a covariance
matrix. Statistical errors on $R_X$, $I_X$, and $T_X$ are obtained
from the fitted models. We adopt distances and errors obtained from
surface brightness fluctuations \citep{Tonry} where available;
otherwise we use LEDA values\footnote{LEDA distances are obtained
from a $B$-band Faber-Jackson relation of the form $M_{B_T}=-6.2 \log
\sigma - 5.9.$} and assume errors of $15\%$.  We adopt a 10\%
uncertainty in $\gamma$ from the galaxy-to-galaxy scatter in the
photon indices fitted to the composite point-source spectra, and a 5\%
uncertainty in $\delta$ from a comparison of values obtained from the
spectral fits and from direct integration of gas-dominated spectra
over the hard and soft bands. We measure the effect on $R_X$ and $I_X$
by repeating the spatial fits with altered values of $\gamma$ and
$\delta$.

Table \ref{xgfp_parameters} lists the base-10 logarithms of $R_X$
(kpc), $I_X$ ($\erg\,\s^{-1}\,\cm^{-2}\,\mbox{$\rm\,arcsec$}^{-2}$)
and $T_X$ ($\kev$) for the final sample of 30 galaxies, and the
corresponding non-zero elements of the covariance matrix.

\begin{deluxetable}{lrrrrrrr}
\tablewidth{0pt}
\tablecaption{Physical parameters for the X-ray gas\label{xgfp_parameters}}
\tablehead{
\colhead{ID} & \colhead{$\log$} & \colhead{$\log$} & \colhead{$\log$}  & \colhead{$C_{RR}$} & \colhead{$C_{II}$} & \colhead{$C_{TT}$} &\colhead{$C_{RI}$}\\
\colhead{} & \colhead{$R_X$} & \colhead{$I_X$} & \colhead{$T_X$}  & \colhead{$\times 10^4$} & \colhead{$\times 10^4$} & \colhead{$\times 10^6$} &\colhead{$\times 10^4$}
}
\startdata
I1459       & $      0.93$ & $    -16.73$ & $    -0.318$ & $     276.3$ & $      48.4$ & $     259.5$ & $     -29.2$ \\
I4296       & $      0.35$ & $    -15.18$ & $    -0.055$ & $      59.3$ & $      32.8$ & $      56.2$ & $     -12.1$ \\
N193       & $      1.34$ & $    -16.79$ & $    -0.114$ & $      44.9$ & $      20.0$ & $      43.3$ & $      -0.6$ \\
N315       & $      0.62$ & $    -15.77$ & $    -0.196$ & $      60.5$ & $      58.1$ & $      75.4$ & $     -12.4$ \\
N533       & $      0.73$ & $    -14.99$ & $    -0.009$ & $      53.1$ & $      90.0$ & $      11.4$ & $     -13.6$ \\
N720       & $      0.96$ & $    -16.46$ & $    -0.247$ & $      48.0$ & $      41.3$ & $      36.9$ & $     -20.0$ \\
N741       & $      0.64$ & $    -15.27$ & $    -0.016$ & $      84.0$ & $      91.5$ & $      72.8$ & $     -30.5$ \\
N1404       & $      0.52$ & $    -15.31$ & $    -0.234$ & $      17.5$ & $      40.7$ & $       7.6$ & $      -1.6$ \\
N1407       & $      0.56$ & $    -15.61$ & $    -0.061$ & $      29.8$ & $      32.5$ & $       6.4$ & $      -1.4$ \\
N1553       & $      1.02$ & $    -17.10$ & $    -0.392$ & $      88.2$ & $    1236.5$ & $     190.2$ & $       2.1$ \\
N2434       & $      1.16$ & $    -17.42$ & $    -0.273$ & $     362.9$ & $      29.7$ & $     621.3$ & $     -79.1$ \\
N3923       & $      0.23$ & $    -15.32$ & $    -0.322$ & $      47.0$ & $      24.1$ & $     217.4$ & $      -7.2$ \\
N4125       & $      1.09$ & $    -16.82$ & $    -0.356$ & $      61.2$ & $      38.0$ & $      57.6$ & $      -6.2$ \\
N4261       & $      0.22$ & $    -15.27$ & $    -0.110$ & $      21.0$ & $      46.2$ & $      56.4$ & $      -5.4$ \\
N4374       & $      0.28$ & $    -15.30$ & $    -0.151$ & $       6.1$ & $      65.0$ & $      15.4$ & $      -0.9$ \\
N4526       & $      0.58$ & $    -16.71$ & $    -0.450$ & $     405.6$ & $      95.5$ & $    1328.8$ & $     -89.2$ \\
N4552       & $      0.01$ & $    -15.20$ & $    -0.245$ & $       9.0$ & $     545.0$ & $      20.1$ & $      -2.2$ \\
N4621       & $      1.29$ & $    -18.05$ & $    -0.629$ & $     456.5$ & $      74.8$ & $    2359.7$ & $    -133.9$ \\
N4636       & $      0.76$ & $    -15.60$ & $    -0.160$ & $      73.1$ & $    1092.3$ & $       0.9$ & $     -98.1$ \\
N4649       & $      0.35$ & $    -15.08$ & $    -0.096$ & $      15.8$ & $      39.4$ & $       2.0$ & $      -7.3$ \\
N4783       & $      1.14$ & $    -16.57$ & $    \phantom{-}0.053$ & $     479.8$ & $     114.8$ & $     878.3$ & $    -105.1$ \\
N5044       & $      1.07$ & $    -15.24$ & $    -0.041$ & $      34.1$ & $      55.8$ & $       1.5$ & $      -3.1$ \\
N5846       & $      0.91$ & $    -15.74$ & $    -0.152$ & $      17.3$ & $      40.0$ & $       4.3$ & $      -0.5$ \\
N6482       & $      1.54$ & $    -16.34$ & $    -0.131$ & $     286.8$ & $     118.1$ & $      21.3$ & $     -67.3$ \\
N7052       & $      1.08$ & $    -16.13$ & $    -0.278$ & $    1323.9$ & $     507.3$ & $     626.0$ & $    -209.1$ \\
N7618       & $      1.59$ & $    -16.34$ & $    -0.095$ & $      73.8$ & $      27.2$ & $      35.4$ & $      -6.8$ \\
\enddata
\end{deluxetable}

\section{The X-Ray Gas Fundamental Plane\label{s.xgfp}}

\begin{figure}
\plotone{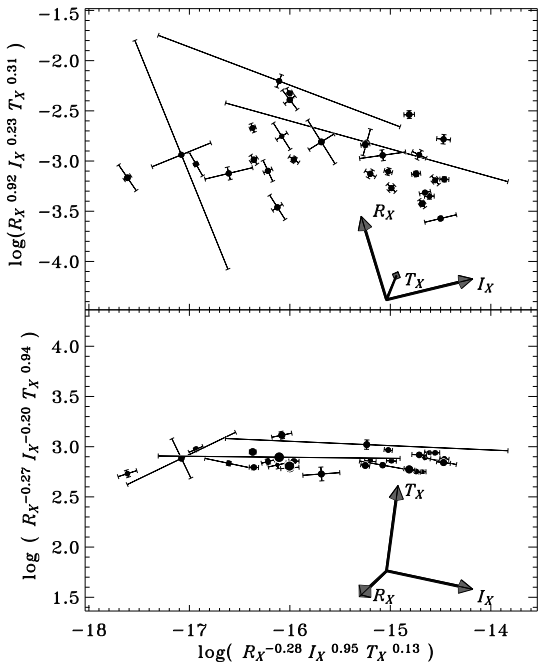}
\caption{Face-on ({\em top\/}) and edge-on ({\em bottom\/}) views of the
X-ray Gas Fundamental Plane. Axis labels indicate eigenvector components. 
Symbol sizes roughly indicate the relative positions into and out of the 
page. Error bars indicate $1\sigma$-projections of the covariance matrices. 
Arrows illustrate the sense of view relative to the fundamental measured 
parameters.
\label{f.plane}}
\end{figure}

The distribution of galaxies in the parameter space
$(\log R_X,\log I_X, \log T_X)$ is nearly planar. This is seen clearly in
Figure \ref{f.plane}, which shows face-on and edge-on views of the XGFP.
We determine the intrinsic distribution in this space by fitting a
probability density in the form of a tilted slab with finite Gaussian width,
taking correlated errors into account.

\begin{figure}
\plotone{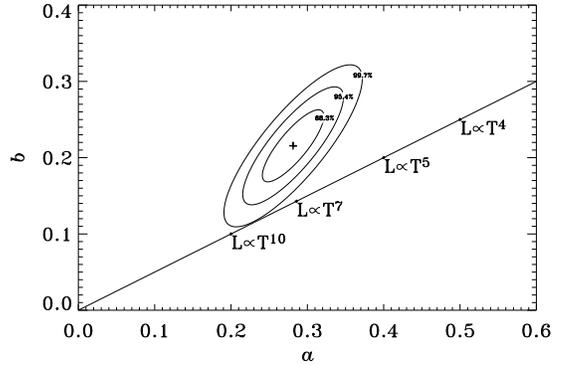}
\caption{Exponents $a$ and $b$, describing the orientation of the XGFP
according to equation (\protect{\ref{f.planefit}}). Cross
and ellipses indicate best-fit values and confidence regions, respectively.
Diagonal line marks combinations of $a$ and $b$ corresponding to pure
luminosity-temperature correlations of the form $L\propto T^n$.
\label{f.coefficients}}
\end{figure}

We can express the XGFP in the form
\beq\label{f.planefit}
T_X \propto R_X^a I_X^b,
\eeq
where $a$ and $b$ determine the orientation of the plane. The best-fit
values are $a=0.28$ and $b=0.22$ (Figure \ref{f.coefficients}). The formal
1-dimensional errors in $a$ and $b$ are $0.045$ and $0.037$, respectively;
but as the figure shows, the errors are correlated. The intrinsic width of
the XGFP is very small, with a value of $0.068 \pm 0.012$ dex, which is
identical to that of the SFP \citep{Sloan_SFP} to within the errors.
The fit is robust and not sensitive to the choice of model for
the surface brightness profile. Excluding the brightest group galaxies or
galaxies with bad single-temperature fits results in $a$ and $b$ values within
the 68.3\% confidence ellipse in
Figure \ref{f.coefficients}.

\section{Discussion\label{s.discussion}}
\subsection{Relation to Known Scaling Laws}

If $a$ and $b$ obeyed the relation $a=2b$, the XGFP would be equivalent to
a simple luminosity--temperature relation of the form
$\lxgas \propto T_X^{2/a}$, indicated by the solid line in
Figure \ref{f.coefficients}. The model fit rules out
a pure $\lxgas$--$T_X$ relation at $>99.7\%$ confidence.
The \lxtx\ relation represents a nearly edge-on view of the XGFP, analogous
to the Faber-Jackson relation \citep{FaberJackson} for the SFP. As in the
optical case, the XGFP accounts for much of the intrinsic scatter ($0.091$ dex)
in the \lxtx\ relation.\footnote{Gas luminosities are obtained by summing
the observed flux in the field of view and using the fitted S\'ersic law to
account for the flux at larger radii.}
For a given $T_X$, galaxies with more extended gas emission
are more luminous than compact objects.

The $L_X$--$T_X$ relation most nearly consistent with the XGFP is
given by $\lxgas \propto T^{8.5}$ (Figure \ref{f.coefficients}). A
naive principal component analysis in the \lxtx\ plane yields a
shallower exponent of $5.9$. \citet{OSullivan} find $L_{X,{\rm tot}}
\propto T_X^{4.8\pm 0.7}$ for their complete sample, and a steeper
exponent of ${5.9\pm1.3}$ when they exclude galaxies with prominent
temperature gradients. However, their X-ray luminosities
include unresolved LMXBs, the removal of
which would steepen the $L_X$--$T_X$ relation.

A comparison can also be made with the $L_X$--$\sigma$ relation for
ellipticals. \citet{Mahdavi} find $L_{X,tot} \propto
\sigma^{10.2^{+4.1}_{-1.6}}$, and predict this relation to steepen to
$\lxgas \propto \sigma^{12 \pm 5}$ if $L_X$ is restricted solely to
the hot gas. Using the temperature-dispersion correlation $\sigma
\propto T_X^{0.56}$ \citep{OSullivan}, we can approximate our closest
$L_X$--$T_X$ relation as $\lxgas \propto \sigma^{15}$, which is
consistent with the earlier result.

\citet{Fujitaa} obtain a result vaguely similar to ours for clusters of
galaxies. They find an X-ray cluster FP connecting core radius, central
density, and mean cluster temperature. Assuming 
a constant value of $\beta = 2/3$ for their surface brightness profile fits,
we can translate
their cluster FP to our parameters, finding $T_X \propto R_X^{0.57} I_X^{0.32}$.
Their relation is significantly inclined to our XGFP, and is close to the relation
$L\propto T^3$. However, the cluster FP deviates from a pure $L$--$T$
relation in a manner similar to that of the XGFP.

\citet{Fukugita} suggest a galaxy fundamental plane with 
mixed X-ray and optical parameters, $L_X$, $R_X$ and $\sigma^2$. Their 
sample consists of 11 galaxies with {\it ROSAT}, {\it Einstein} or 
{\it ASCA} observations. We do not reproduce their result with our larger
{\it Chandra\/} sample.
The reason for this discrepancy may be that their X-ray luminosities are
corrected neither for light outside the field of view nor for the 
contribution of point sources.

\subsection{Independence of the XGFP and the SFP}

The virial theorem connects a system's total mass $M$ with its characteristic
radius $R_M$ and dispersion $\sigma_M$. This relation
produces an observable SFP because mass maps to luminosity by way of the
mass-to-light ratio, and the stellar $R_e$ and $\sigma$ are surrogates for
$R_M$ and $\sigma_M$. For the XGFP to be another manifestation of the virial
theorem, one would require similar mappings from mass parameters into X-ray
observables. One example is hydrostatic equilibrium, which 
links $R_X$ and $T_X$ for a given potential. Others might connect
gas mass to total mass or to a measure of gas retainability such as
$\sigma^2/T_X$. However, these relations would be identifiable
in correlations between optical and X-ray parameters. Only the known
$T_X$--$\sigma$ relation
\citep[e.g.][]{OSullivan} and an additional, very weak $T_X$--$R_e$ relation
are supported by the data. X-ray gas masses are uncorrelated with $M_B$ and
other optical properties.

If the SFP and XGFP were linked, then each plane would represent a
projection of a higher dimensional, more fundamental hyperplane into the
corresponding 3-parameter subspace. We test this hypothesis by analyzing the
($\log R_X$, $\log I_X$, $\log T_X$, $\log R_e$, $\log I_e$, $\log \sigma^2$)
space with principal component analysis (PCA). In this space
our sample is reduced to 25 objects with reliable X-ray and optical
parameters. We
define the 3-vectors $\vv{n_X}$ and $\vv{n_O}$ to be the normals to the
fundamental planes in X-ray and optical parameters, respectively. From the
SFP we have 
$\vv{n_O}=[0.69, 0.51, -0.51]$ \citep{Sloan_SFP} 
and from the XGFP,
$\vv{n_X}=[0.26, 0.21, -0.94]$. If the planes are completely independent, the
eigenvectors corresponding to the two smallest eigenvalues will have the form
$N_1=[\mu\,\vv{n_X}, \sqrt{1-\mu^2}\,\vv{n_O}]$ and
$N_2=[\sqrt{1-\mu^2}\,\vv{n_X}, -\mu\,\vv{n_O}]$.
Here, $N_1$ and $N_2$ are 6-vectors,
and $\mu$ can be any number between $-1$ and 1, depending on the relative
scatter in the two planes. We obtain
$N_1 = [-0.21, -0.16, 0.84, 0.20, 0.15, -0.39]$ and
$N_2 = [0.01, 0.03, -0.38, 0.79, 0.38, -0.28]$.
This result is not far from the above prediction 
with $\mu = -0.40$. Furthermore, the scatter about both planes is not reduced
by going to higher dimensions. Both results point to, at most, weak coupling
of the subspaces. 

Optical and X-ray parameters are known to be coupled through the
observed $T_X$--$\sigma^2$ relation. We test whether this coupling could
have a measurable effect on the eigenvectors, using
Monte Carlo simulations of 25-object samples following the fitted SFP
and XGFP relations. We take the SFP and XGFP to be independent, except for
an intrinsic linear correlation between $\log T_X$ and $\log \sigma^2$ with
a variable amount of scatter. We consider three coupling strengths:
none, weak, and strong. Weak coupling reproduces the observed width
($0.10$ dex)
of the $T_X$--$\sigma^2$ relation and, by way of the SFP, the somewhat
weaker $T_X$--$R_e$ relation; strong coupling reduces the $T_X$--$\sigma^2$
width by a factor of 3. We perform PCA on the simulated 6-d data and
measure the alignment of the best 2 eigenvectors with those obtained from the
real data. All 3 cases reproduce the eigenvectors to within the Poisson noise,
showing that the observed $T_X$--$\sigma^2$ relation does not change the
relation between a decoupled XGFP and SFP at a level that can be resolved with
a 25-object sample.

The data are thus consistent with the XGFP and SFP being almost
completely independent. The XGFP cannot be understood as a simple
consequence of the virial theorem or hydrostatic equilibrium. Instead,
the XGFP represents a new constraint on the hydrodynamic evolution of
elliptical galaxies.

\acknowledgments
We are grateful to Daniel Wang and Zhiyuan Li for helpful discussions.
We have made use of the HyperLEDA database (http://leda.univ-lyon1.fr).
Support for this work was provided by the National Aeronautics and Space
Administration (NASA) through Chandra Awards G01-2094X and AR3-4011X,
issued by the {\em Chandra X-Ray Observatory Center}, which is operated by the
Smithsonian Astrophysical Observatory for and on behalf of NASA under
contract NAS8-39073.

\end{document}